\documentclass[letterpaper, 10 pt, conference]{ieeeconf}  

\IEEEoverridecommandlockouts                              
\usepackage{graphicx}
\usepackage{amssymb}
\usepackage[flushleft]{threeparttable}
\usepackage{amsmath}
\usepackage{lineno}
\usepackage{color,soul} 
\soulregister\cite7
\soulregister\ref7
\usepackage{flushend}
\allowdisplaybreaks

\title{\LARGE \bf
On Wind Speed Sensor Configurations and Altitude Control \\ in Airborne Wind Energy Systems
}

\author{Laurel N. Dunn$^{1,3}$, Christopher Vermillion$^{2}$, Fotini K. Chow$^{1}$, and Scott J. Moura$^{1}$
\thanks{$^{1}$Department of Civil \& Environmental Engineering, University of California Berkeley}
\thanks{$^{2}$Department of Mechanical Engineering, North Carolina State University}
\thanks{$^{3}$Corresponding author: {\tt\small lndunn@berkeley.edu}}%
}

\begin{document}

\maketitle
\thispagestyle{empty}
\pagestyle{empty}

\begin{abstract}
Real-time altitude control of airborne wind energy (AWE) systems can improve performance by allowing turbines to track favorable wind speeds across a range of operating altitudes.
The current work explores the performance implications of deploying an AWE system with sensor configurations that provide different amounts of data to characterize wind speed profiles.
We examine various control objectives that balance trade-offs between exploration and exploitation, and use a persistence model to generate a probabilistic wind speed forecast to inform control decisions.
We assess system performance by comparing power production against baselines such as omniscient control and stationary flight.
We show that with few sensors, control strategies that reward exploration are favored.
We also show that with comprehensive sensing, the implications of choosing a sub-optimal control strategy decrease.
This work informs and motivates the need for future research exploring online learning algorithms to characterize vertical wind speed profiles.
\end{abstract}

\section{Introduction and Motivation}
\label{sec:intro}
Airborne wind energy (AWE) turbines are an emerging wind generation technology. 
AWE systems differ from conventional turbines in that they are attached to the ground by an adjustable tether, rather than by a fixed tower.
This tethered design reduces capital costs and makes it possible to achieve a higher capacity factor.
We refer readers to \cite{Diehl2013} for an in-depth description of airborne wind energy.
Information about modern AWE systems can be found at \cite{altaeros2018,skysails2018,ampyx2018}.

Three characteristics of AWE systems contribute to a high capacity factor compared with conventional turbines.
First, AWE systems can reach higher altitude winds, which tend to be stronger and more consistent than surface-level winds \cite{archer2018}.
Second, AWE systems can use crosswind flight patterns to increase the apparent wind speed \cite{loyd1980}, leading to more power output than would be achieved with stationary flight \cite{fagiano2014, vermillion2018}.
Finally, by adjusting the operating altitude in real-time, AWE systems can vertically track favorable wind speeds in a spatially and temporally varying wind field \cite{archer2014}.
The current work focuses on understanding how sensor placement impacts the performance of an AWE system with various altitude control schemes.

Currently, research on AWE altitude control systems focus on control schemes that find the optimal operating altitude \cite{bafandeh2016} and are robust to uncertainty \cite{baheri2017}.
For example, \cite{baheri2017} borrow control objectives from Bayesian optimization to balance exploration and exploitation.
More specifically, these objectives balance the trade off between capitalizing on known wind resources (exploitation) and collecting observations at altitudes where wind speed estimates are uncertain (exploration), given that wind speed profiles are only partially observable and that altitude adjustments are costly to make.
Over time, control decisions favoring exploration can reduce uncertainty in wind speed models, though doing so may come at a cost to near-term performance.
We refer readers to \cite{brochu2010} for information on Bayesian optimization and a more detailed discussion of the trade-off between exploration and exploitation.

In the context of AWE system altitude control, this trade off is directly related to uncertainty in wind speed estimates that inform control decisions.
A survey of state-of-the-art algorithms for forecasting wind power production is given in \cite{soman2010}.
Many of the algorithms cited require large training data sets.
Training such algorithms may not be possible for AWE systems which typically rely on sparse data streams collected online.
Observations are sparse because wind speeds are recorded by a single sensor that tracks (vertically) with the operating altitude of the turbine.
A simpler persistence model is also shown to perform quite well, and is recommended as a benchmark for evaluating the performance of more complex wind forecasting algorithms.

Aside from using complex forecasting algorithms, uncertainty in wind speed can be reduced by increasing the spatial coverage of wind speed sensors. 
For example, if wind speeds were recorded continuously at all altitudes, a simple forecasting model (e.g. the persistence model) may be able to outperform a system that uses a more sophisticated forecasting model trained on single sensor data.
This possibility motivates a comparative analysis of different sensor configurations and forecasting methods for AWE altitude control, an issue currently unexplored in the literature.


The main contribution of the current work is to develop a framework for evaluating performance gains achievable using different wind speed sensor configurations.
We demonstrate this framework with a case study of a particular wind field.
We use the control objectives proposed in \cite{baheri2017} to determine the optimal altitude trajectory.
Questions related to the performance gains from coupling sensor data with different statistical forecasts are reserved for future research.

This paper is organized as follows. 
Section \ref{sec:methods} provides methodological details.
Section \ref{sec:observability} outlines the sensor configurations. 
Section \ref{sec:forecast_methods} details the forecasting methods.
Section \ref{sec:control_methods} formulates the altitude control objectives. Section \ref{sec:benchmarks} describes the comparative analysis benchmarks and metrics. 
Section \ref{sec:results} provides the results and discussion. Finally, Section \ref{sec:conclusions} summarizes the paper's main conclusions.


\section{Methods}
\label{sec:methods}
\begin{figure}[t]
    \centering
    \includegraphics[width=\linewidth, trim=0 4mm 0 4mm,clip]{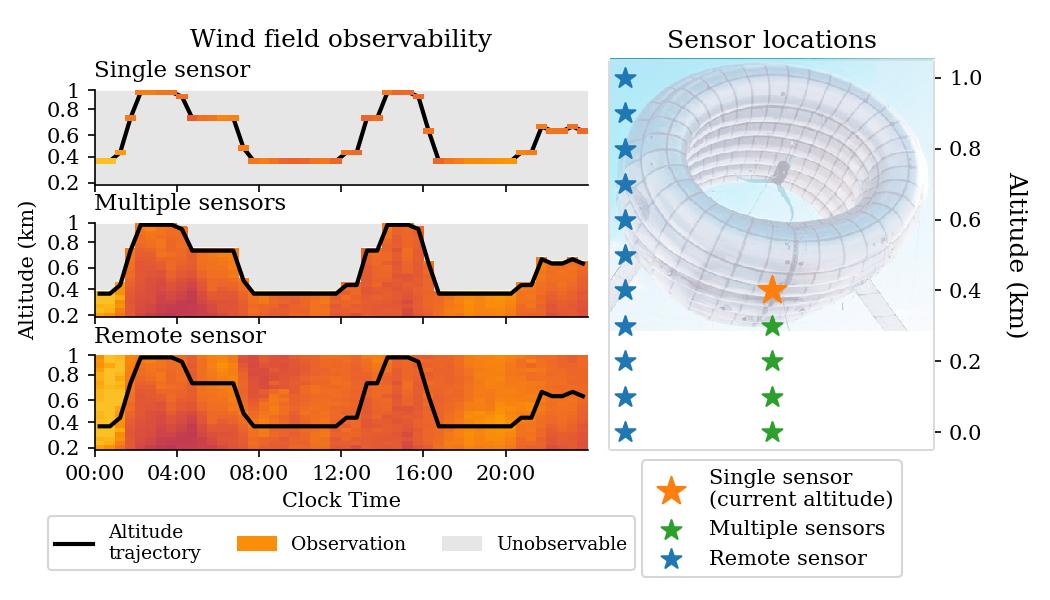}
    \caption{Schematic diagram showing wind field measurements recorded by each of three sensor configurations. Right panel shows measurement locations for single, multiple, and remote sensor configurations. Left panel shows the altitude (y-axis) and magnitude (color scale) of the measurements that would be recorded by each sensor configuration given some trajectory (black line) of altitudes with respect to time (x-axis).}
    \label{fig:turbine}
\end{figure}

In this work we simulate the altitude trajectory of a buoyant airborne turbine (pictured in Fig. \ref{fig:turbine}) in a spatially and temporally varying wind field.
The simulation relies on wind speed data recorded by a 915-MHz wind profiler between July 1, 2014 and August 31, 2014 at Cape Henlopen State Park in Lewes, DE \cite{archer2018}.
Wind speed data are measured every 50 meters in 30 minute intervals. We use the same spatial and temporal discretization in simulation.

We examine single-sensor, multiple-sensor and remote sensor configurations. 
We track the wind speed measurements that would be recorded online given a particular sensor configuration and the altitude trajectory followed up to a particular point in time.
Observations are used to train a persistence model that generates a probabilistic wind speed forecast.
We use three different control objectives to determine the optimal altitude trajectory given the current wind speed forecast.

Altitude trajectories are generated for a range of scenarios, each of which uses:
\begin{enumerate}
    \item One of three control objectives 
    \item One of three sensor configurations
    \item A persistence forecast
\end{enumerate}
The current work examines differences in power production with each sensor configuration.
We build on the control objectives presented in \cite{baheri2017} to do so, and leave it to future work to explore the performance implications of using more sophisticated wind forecasting methods.


\section{Sensor configurations}
\label{sec:observability}

\subsection{Single-sensor configuration} 
AWE systems are typically designed with a single anemometer to measure the wind speed at the current operating altitude.
The vertical position of these measurements changes when altitude adjustments are made.

\subsection{Multiple sensor configuration}
A novel sensor design would be to record wind speed measurements along the length of the tether.
The system is still considered partially observable as wind speeds are not measured above the hub height.
However, flying the turbine at the highest altitude within operating bounds would provide a complete wind speed profile.

The focus of the current work is to assess the performance implications rather than to explore sensor technologies themselves.
However, we have identified several technologies that could collect these measurements.
One option is to affix anemometers in regular intervals along the tether, which provides reliable wind speed measurements at altitudes below the AWE system but requires a winch system to house the anemometers upon spool-in.
Another alternative is to attach telltales to the tether and use image processing to estimate wind speeds from the angle of the telltale relative to the tether.
Finally, one could potentially compute local wind velocities based on the catenary geometry of the tether, though to do so would require a detailed characterization of the tether's structural and aerodynamic properties. 
Further research is needed to assess the technical and economic feasibility of specific solutions for collecting these measurements. 


\subsection{Remote sensor configuration}
The third configuration relies on a remote sensor to record wind speed measurements in discrete intervals across a wide range of altitudes simultaneously.
For example, the wind data we use to inform our simulation was collected using a vertical profiler that measures wind speeds in 50 meter increments.
With remote sensing wind speed profiles are fully observable at each time step.
Because the same measurements are recorded regardless of the current position of the turbine, this configuration decouples data acquisition from altitude control.
The implication is that exploring the wind field for data acquisition is not necessary, and that control decisions can focus instead on exploitation of known wind resources.

\section{Persistence Forecast}
\label{sec:forecast_methods}
We use a persistence model adapted from \cite{soman2010} to generate a probabilistic wind speed forecast.
We extend the model to extrapolate wind speeds into the spatial domain, and to characterize uncertainty. 
The persistence model is based on the premise that wind speeds change very little (or not at all) from one time step to the next.
The spatial extension of this model presumes that there is little (or no) change vertically either.
Based on this premise, the forecast mean $\mu$ at altitude $h$ and $s$ time steps into the future is given by:

\begin{equation}
\label{eq:persistenceMean}
\mu_{h,t+s} = X_t(h')
\end{equation}
where $X_t(h)$ describes the wind speed observation recorded at time $t$ and altitude $h$.
We define $h'$ to be the measurement altitude in $X_t$ that is either equal to the forecast altitude $h$, or is nearest to it.
For example, with multiple sensors, $h'$ would be equal to $h$ for altitudes below the current operating altitude, and would equal the current altitude (i.e., the highest observable altitude) otherwise.
The fundamental assumption is that the measurement $X_t(h')$ recorded at ($h',t$) persists across all unobserved times and altitudes.

We characterize uncertainty by estimating how erroneous the assumption of persistence could be based on past observations.
We define $\mathbf{\Delta}$ to be a matrix composed of column vectors $\Delta X/\Delta h$ and $\Delta X/\Delta t$ describing the finite differences between measurements in $X$ in space and time.
Based on an exploratory analysis of the data, we characterize $\mathbf{\Delta}$ as a joint Gaussian distribution with zero mean.

Next, we define the vector $\mathbf{d}$ describing the distance in space ($h-h'$) and in time ($s$) between the current observation ($X_t(h')$) and the forecast.
We can then characterize forecast uncertainty ($\sigma^2$) as:

\begin{equation}
\label{eq:persistenceVariance}
\begin{aligned}
\sigma^2_{h,t+s} &= \Sigma \left( \mathbf{d} \mathbf{\Delta}^T \right) = \mathbf{d} \Sigma(\mathbf{\Delta}) \mathbf{d}^T \\
\end{aligned}
\end{equation}

\noindent
Here $\Sigma(\mathbf{\Delta})$, for example, is the covariance of $\mathbf{\Delta}$.




Where forecast uncertainty is high, we truncate the distribution between 0 and 17 m/s to ensure that high uncertainty does not lead to unrealistic predictions.
We find that 98\% of all observations in the data are within these bounds.
We assume that the underlying distribution of $\mathbf{\Delta}$ is stationary, though characterizing non-stationarity in wind speed dynamics presents an interesting opportunity for future work.


\section{Control Methods}
\label{sec:control_methods}

We describe altitude control using the following simple integrator dynamics:
\begin{equation}
\label{eq:dynamics}
h(t+1) = h(t) + u(t)
\end{equation}
where $h(t)$ is the altitude at discrete time index $t$, and $u(t)$ is the controlled altitude adjustment. 

The control objective at a given time $t$ is to select the trajectory of optimal future operating altitudes $\{h^*(t+1),h^*(t+2),\cdots\}$ and controlled altitude adjustments in $\{u^*(t),u^*(t+1),\cdots\}$ that maximize some objective function, given the wind speed forecast $V$.
We express this objective $J$ mathematically as:
\begin{equation}
\label{eq:objective}
\begin{aligned}
\underset{h(t),u(t)}{\max}\,\,J &= \sum_{s=t}^{t+T} g(h(s), u(s),  V_{\forall h,s})\\
\end{aligned}
\end{equation}

\noindent
Here $V_{\forall h,s}$ to refers collectively to the random wind speed forecasts for all candidate altitudes $h$ at time $s$, and $T$ is the planning horizon.


The problem is constrained such that operating altitudes are bounded to within $h_{\min}$ and $h_{\max}$, and the rate of change in altitude is below $r_{\max}$.
\begin{equation}
\label{eq:constraints}
\begin{aligned}
h_{\min} \le h(t) \le h_{\max} &  \\
\left|u(t)\right| \le r_{max} \\
\end{aligned}
\end{equation}
Table I provides numerical values for operating constraints.

This formulation uses model predictive control to optimize the control and state trajectories over the upcoming $T$ time-steps given the current state $h(t)$, dynamical model \eqref{eq:dynamics} and wind speed forecasts for all altitudes. 
Only the first control action $u^*(t)$ is physically implemented, and the process is repeated using the measured state in the next time step. 

We use a planning horizon of 90 minutes (or three 30-minute time steps). 
Given the values listed in Table I, the planning horizon is sufficient for the turbine to travel between any two operating altitudes. 
We use dynamic programming to solve for the optimal trajectory at each time step.

\begin{table}[t]
\begin{center}
\label{tab:constants}
\caption{List of model parameters and values.}
\begin{tabular}{|c|l|}
\hline\textbf{Variable} & \textbf{Value} \\
\hline \hline
    $h_{min}$ & 0.15 km \\
    $h_{max}$ & 1.0 km \\
    $r_{max}$ & $0.01$ km/min \\
    $v_{r}$ & 12 m/s \\
    $\Delta t$ & $30$ min \\
    $T$ & $90$ min \\
    $k_1$ & 0.0579 $\text{kW s}^3/\text{m}^3$ \\
    $k_2$ & 0.09 $\text{kW s}^2/\text{m}^2$ \\
    $k_3$ & 1.08 $\text{kW s}^2/\text{m}^2\cdot \text{km}$ \\
\hline
\end{tabular}
\end{center}
\end{table}



We examine three formulations of $g(h,u, V)$ that aim to maximize power production.
These formulations differ in how they account for uncertainty in power production (which stems from uncertainty in wind speed forecasts).
Although the objective function differs for each formulation, the long-term goal is always to maximize power production.

Equation \eqref{eq:power_production} lists the system of equations used to calculate power production, as described in \cite{bafandeh2016}.
Here power production $p(u(t),v)$ is a function of the altitude adjustment during some time interval $u(t)$ and the true wind speed $v$.

\begin{eqnarray}
\begin{aligned}
\label{eq:power_production}
p(u,v) &= p_{1}-p_{2}-p_3 \\
p_1 &= k_1 \cdot \min\{v_r, v\}^3 \\
p_2 &= k_2 v^2 \\
p_3 &= k_3 v^2 \cdot \left| u \right| \\
\end{aligned}
\end{eqnarray}

In words, the total power production $p(u,v)$ is the difference between the amount of energy the turbine generates ($p_1$) and the amount of energy required to maintain ($p_2$) and to adjust ($p_3$) the operating altitude.
The rated wind speed of the turbine is given by $v_r$, and $k_1$, $k_2$ and $k_3$ describe lumped parameters representing the mechanical and aerodynamic properties of the system.
Numeric values for these constants are provided in Table I.

We highlight that $p_1$ is maximized when $v$ is equal to $v_r$.
However, $p_2$ and $p_3$ continue to increase at wind speeds greater than $v_r$.
The implication is that $p(u,v)$ increases as $v$ approaches $v_r$, but decreases if $v$ increases beyond $v_r$.

The wind speed $v$ can represent either a measurement reported in the data, or some realization of the wind speed forecast.
In order words, if $V_{h,t}$ is a random variable describing the wind forecast at altitude $h$ and time $t$, then we can use the function $p(\cdot)$ to derive a probabilistic forecast of power production, denoted by $P_{h,t}$.
Though $P_{h,t}$ is not explicitly related to $h$ or $t$, it is implicitly related by the fact that the wind forecast changes with respect to both quantities.

In the following sections we describe three candidate formulations of $g(h,u,V)$.
Though the specific objective functions are different, the aim of all three formulations is to maximize overall power production.
The objective functions are borrowed from Bayesian optimization, and their application to real-time control of AWE systems is motivated in \cite{baheri2017}.

Though the formulations we use are conceptually the same as the control objectives presented in \cite{baheri2017}, we have adapted them in two important ways.
First, we optimize over a finite planning horizon extending $T$ time steps into the future. 
Second, we use a probabilistic wind speed forecast to compute a probability distribution of power production.
Since power production is a nonlinear function of wind speed, the power production is not Gaussian, and generally does not follow a parametric distribution.
As shown below, our approach handles non-parametric distributions directly, and does not require parametric approximations.

\subsection{Maximize Expected Energy}
The first control strategy chooses the altitude trajectory that maximizes expected power production across time steps within the planning horizon. This can be viewed as an exploitative control approach, in the sense that no reward is explicitly provided to explore portions of the state-space where uncertainty is high. Instead, the goal is to directly maximize expected power production.

In continuous form, the expected power at time $t$ is given by  \eqref{eq:expectation_continuous}, where $f_{V_{h,t}}(v)$ is the probability density function (PDF) of wind speed forecast random variable $V_{h,t}$.

\begin{equation}
\label{eq:expectation_continuous}
\begin{aligned}
     \mathbb{E}[P_{h,t}|u,V_{h,t}] &= \int_{0}^{\infty} f_{V_{h,t}}(v) p(u,v) dv \\
\end{aligned}
\end{equation}

\noindent

To accommodate non-parametric wind speed forecasts with no closed form solution to \eqref{eq:expectation_continuous}, we use the discrete approximation given by
\begin{equation}
\label{eq:expectation_discrete}
\begin{aligned}
     \mathbb{E}[P_{h,t}|u,V_{h,t}] &\approx \frac{1}{n}  \sum_{q=1}^{n}  p(u,v_{q/n}) \\
     v_{q/n} &= Q_{V_{h,t}}(q/n) \\
     Q_{V_{h,t}}(q/n) &:= \Pr(V_{h,t} \le v) = q/n \\
\end{aligned}
\end{equation}

\noindent
Here $Q_{Y}(q/n)$ is the inverse cumulative density function (CDF) of some random variable $Y$ evaluated at quantile $q$ for a specified number of quantile bins $n$.
For example, $Q_{Y}(q/n)$ evaluates to the $q^{th}$ quartile of $Y$ when $n$ is 4, or to the $q^{th}$ percentile when $n$ is 100.
We set $n$ equal to 100.



\subsection{Maximize Upper Confidence Bound (UCB)}
The second control strategy chooses the altitude trajectory that maximizes performance under an optimistic realization of the forecast. 
This is also known as quantile optimization.
For example, one might maximize the $90^{th}$ percentile of $P_{h,t}$.
Equation \eqref{eq:upper_confidence_bound} expresses this mathematically for an arbitrary probability threshold $\alpha>0.5$ and usually near $1$.

\begin{equation}
\label{eq:upper_confidence_bound}
g(h,u,V) = Q_{P_{h,t}|u,V_{h,t}}(\alpha | u,v)
\end{equation}
where $Q_{P_{h,t}|u,V_{h,t}}$ describes the inverse conditional CDF of the random variable $P_{h,t}$ conditioned on $u=u$ and $V_{h,t}=v$.

This approach rewards trajectories that explore altitudes where uncertainty may be large yet potentially yields high power production.
Tuning the upper confidence bound $\alpha$ adjusts the degree to which exploration is rewarded.
The drawback is that when an overly optimistic control objective is used (i.e., if $\alpha$ is close to 1), there is a high risk that the observed power production will be much lower than the value used to inform a control decision.

Algorithms favoring exploration may under-perform relative to purely exploitative methods, except insofar as they reward acquisition of new data that reduces long-run uncertainty in the wind speed forecasts.
As uncertainty bounds become narrow, the difference in power production between control objectives favoring exploration and exploitation also decreases.


\subsection{Maximize Probability of Improvement}

The last control strategy selects the altitude trajectory with the highest probability of improving performance relative to maintaining the current altitude.
We calculate the probability of improvement by taking the log probability that power production for a particular trajectory will exceed power production if the altitude were to remain fixed at current altitude $h$.
Mathematically, this is expressed in \eqref{eq:probability_improvement}.
\begin{equation}
\begin{aligned}
\label{eq:probability_improvement}
    g(h,u,V) = \log \Pr \left( P_{h+u,t} > p(0,v) \right)
\end{aligned}
\end{equation}
where $P_{h+u,t} = p(u,V_{h,t})$ is a random variable describing power generation at altitude $h+u$, and $p(0,v)$ is the current power output (i.e. no altitude adjustment $u=0$ and assuming wind speed $v$ stays constant).

\section{Performance Metrics and Baseline Scenarios}
\label{sec:benchmarks}
In the current section we define metrics and benchmarks to evaluate the performance of each sensor configuration and control strategy.
We calculate these metrics and benchmarks by simulating the altitude trajectory over the course of three months.

\subsection{Performance metrics}
The most fundamental metric we use to evaluate performance is average power production (in kW).
Though power production could be compared against the nameplate capacity (in this case 100 kW), this is not a practical target as it is not physically possible to achieve that level of performance.
A more practical target would need to account for the physics of the simulation environment, including variations in wind speed and the energy required to adjust and maintain altitude.
The omniscient baseline (described below) provides just such a target.
In addition to reporting power production, we express performance as a ratio (between 0 and 1) of the energy harvested in a particular scenario and the energy harvested in the omniscient baseline.
We refer to this quantity as the ``actualized power ratio'' (as in Figs. \ref{fig:ucb-thresholds} and \ref{fig:barchart}).

\subsection{Performance baselines}
\subsubsection{Omniscient baseline}
The omniscient baseline is obtained by simulating the altitude trajectory the AWE would follow if perfect information were available to inform control decisions.
The result provides an upper limit on power production, given the characteristics of the wind field and the specified operating constraints.

\subsubsection{Fixed altitude baseline}
We also compare our results against baselines where the altitude is fixed for the entire of the simulation at the altitudes that would achieve the highest ($h_{\text{best}}$) and lowest power production ($h_{\text{worst}}$).
Though these provide a useful basis for comparison, we note that it is not possible to know $h_{\text{best}}$ or $h_{\text{worst}}$ without omniscient information about the wind speeds at each altitude.

Unlike the omniscient baseline, fixed altitude trajectories do not bound system performance. 
Instead, they provide a benchmark against a na{\"i}ve, but possibly effective, control strategy.
A real-time control scheme can under-perform relative to a fixed altitude trajectory if control decisions are made based on sufficiently erroneous wind speed forecasts, or if the power production observed at the new operating altitude does not compensate for the energy expended in making the altitude adjustment.


\section{Results and Discussion}
\label{sec:results}
Here we summarize the performance of an AWE system evaluated in simulation for nine scenarios.
What differentiates each scenario is the specific combination of sensor configuration and control scheme used to inform altitude adjustments.
A persistence forecast is trained on observational data collected in simulation, given the sensor configuration and altitude trajectory up to that point.
We examine a range of values for $\alpha$ in the upper confidence bound (UCB) control and report results for the value that achieves the highest performance in each sensor configuration.
We compare the performance in each scenario against the omniscient and fixed altitude baselines.

\begin{figure}[t]
    \centering
    \includegraphics[trim = 9mm 6mm 7mm 5mm, clip, width=\linewidth]{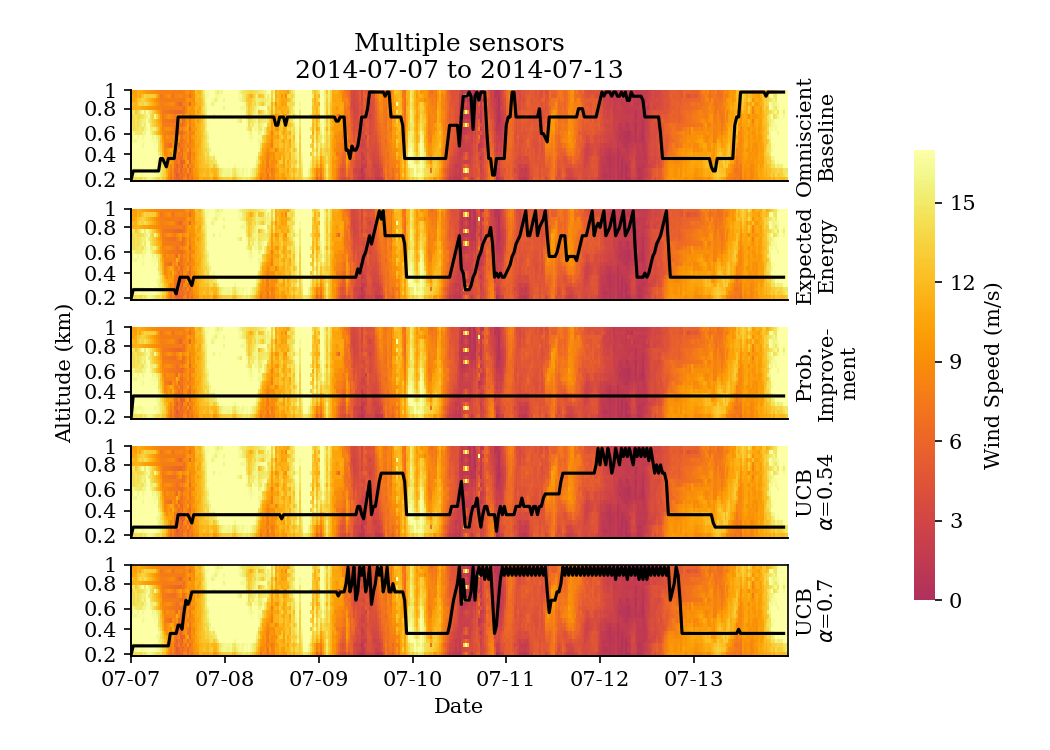}
    \caption{Altitude trajectory of AWE turbine (black line) over one week (July 7-13, 2014) for control scenarios indicated on the right. We include UCB control scenarios where $\alpha=0.54$ (the optimal), and $\alpha=0.7$ (included for illustrative purposes). The color scale indicates the wind speed at each altitude (y-axis) with respect to time (x-axis).}
    \label{fig:heatmap}
\end{figure}

Fig. \ref{fig:heatmap} shows the altitude trajectory over one week in August for the omniscient baseline and four control schemes in the multiple-sensor scenarios.
Commenting on the similarities and differences between trajectories highlights the merits (and pitfalls) of using a particular control scheme with each sensor configuration.

\emph{Observation 1:} Though altitude trajectories follow very different patterns at times when the wind speed is low (e.g., July 11-13), they all follow a relatively fixed course when wind speeds are high (e.g., July 8-9).
The reason for this is that $p_1$ is constant for wind speeds in excess of $v_r$, while $p_2$ and $p_3$ continue to increase.
The incentive to explore is only in place if the potential increase in power production exceeds the cost of making altitude adjustments.
When the wind speeds are near or in excess of $v_r$, a fixed altitude is favored because exploration comes at a relatively high cost without the possibility of increasing power production.

\emph{Observation 2:} Maximizing the probability of improvement leads to a fixed altitude trajectory in both the single- and multiple-sensor cases.
The reason for this is that wind speed forecasts are centered around the current observation.
In other words, the forecast estimates that exploring some unobserved altitude is equally likely to reduce performance as to improve performance.
Thus the probability of improvement is only 50\%, and control decisions favor maintaining a constant altitude to avoid power loss from making altitude adjustments.

Though the objective to maximize expected energy is also centered about the current observation, control decisions in that case are informed not only by the probability but also the magnitude of improvement potential.
Since the magnitude of improvement scales with $v^3$, the distribution of $p$ is skewed to the right and there is some incentive to explore.

\emph{Observation 3:} In the multi-sensor case, when the objective is to maximize the upper confidence bound (UCB), trajectories tend towards higher altitudes rather than lower altitudes.
This happens because the uncertainty is greater at unobserved altitudes above the current hub height than at altitudes where current measurements are available.
This high uncertainty creates a strong incentive to explore higher altitudes.
However, once the highest altitude is reached, the system becomes completely observable and uncertainty is equal at all altitudes, so exploitation is favored.

\begin{figure}
    \centering
    \includegraphics[trim = 0mm 4mm 0mm 8mm, clip, width=\linewidth]{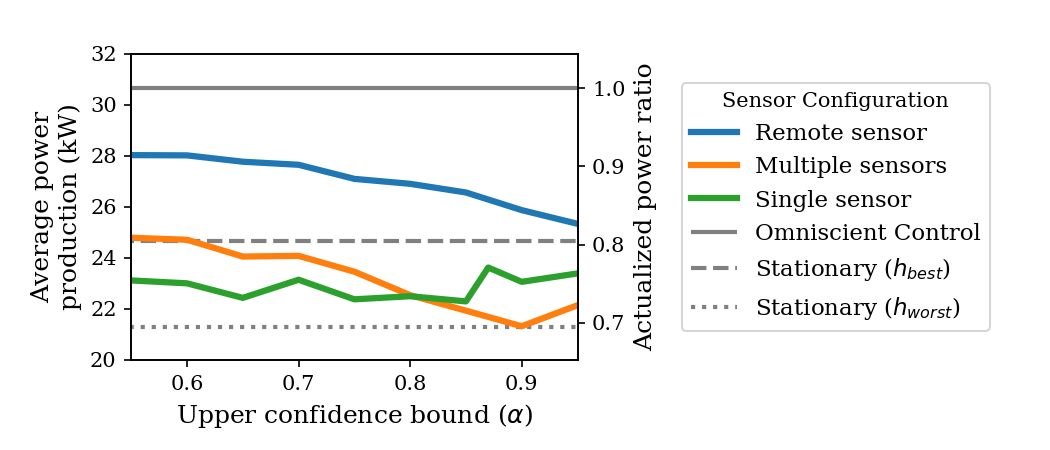}
    \caption{Average power production (y-axis) for upper confidence bound control scenarios, as a function of confidence level $\alpha$ (x-axis).}
    \label{fig:ucb-thresholds}
\end{figure}

At this point the trajectory will tend downwards if the best wind resource is below the uppermost altitude.
Decreasing the hub height also makes the system only partially observable, reinstating the reward to explore higher altitudes.
This process repeats, causing the oscillations observed in the lowermost panel in Fig. \ref{fig:heatmap}. 
These oscillations come at a high energy cost and do not necessarily lead to gains in overall performance.
We examine how energy is allocated when $\alpha$ is set to 0.7, and compare it against energy allocation when the optimal value (0.54) is used.
Although the higher incentive to explore leads to a 4\% increase in power production ($p_1$), the system expends twice as much energy on altitude adjustments ($p_3$).
The additional energy cost leads to a 3\% reduction in overall performance.
Fig. \ref{fig:ucb-thresholds} shows that performance tends to decrease as $\alpha$ increases.

\begin{figure}
    \centering
    \includegraphics[trim = 0mm 0mm 0mm 8mm, clip, width=\linewidth]{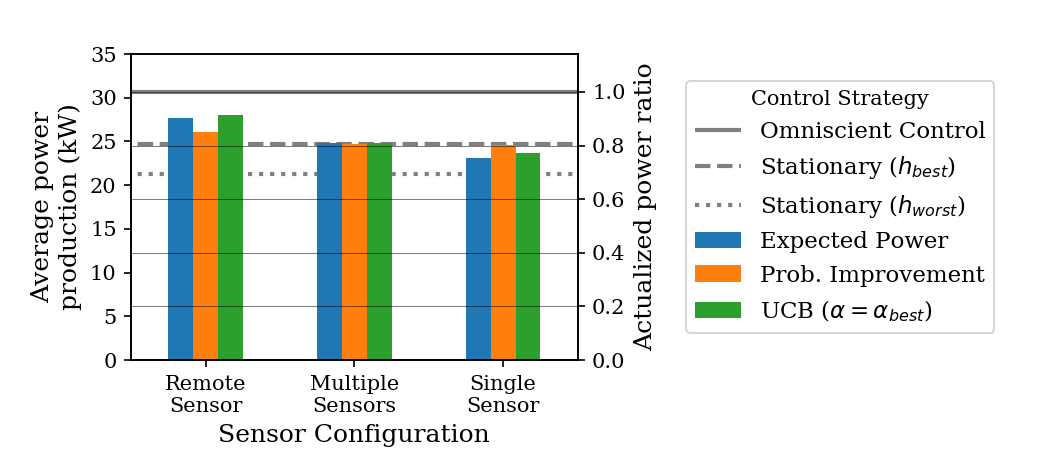}
    \caption{Average power production for each control scheme \& sensor configuration. Horizontal lines denote omniscient baseline (solid line), and the best (dashed line) and worst (dotted line) fixed altitude trajectories. Performance is measured in terms of average power production (left y-axis) and in terms of the ``actualized power ratio'' between power production and the omniscient baseline (right y-axis).}
    \label{fig:barchart}
\end{figure}

Fig. \ref{fig:barchart} summarizes overall power production across the three-month simulation in all nine scenarios, and compares them against baselines presented in Section \ref{sec:benchmarks}.

Our results show that the remote sensor improves performance by 11\% compared with deploying the system with multiple sensors, and by 15\% compared with deploying a system with only a single sensor.
Fig. \ref{fig:ucb-thresholds} shows that although performance declines as $\alpha$ increases, the remote sensor consistently outperforms the other two sensor configurations.
In all scenarios the multiple and single-sensor cases perform about on par with the optimal altitude ($h_{best}$) for stationary control.
In practice, however, it is not possible to know $h_{best}$ in advance.

These results are based on the highest performing control strategy for each sensor configuration.
However, the optimal value for $\alpha$ is not known \textit{a priori}, and likely depends on many factors such as the spacing of sensors and the characteristics of the wind field.
Fig. \ref{fig:ucb-thresholds} shows that the consequences of choosing a sub-optimal control strategy are particularly severe in the multiple sensor configuration.

Finally, Fig. \ref{fig:ucb-thresholds} shows that heavily rewarding exploration may slightly improve performance in the single-sensor configuration, but actually decreases performance in the multi-sensor configuration.
This decline in performance is due to the altitude oscillations discussed in Observation 3.

\section{Conclusions and Future Work}
\label{sec:conclusions}
In this work we report on differences in AWE performance achieved in simulation using various different control strategies and wind speed sensor configurations.
The key difference between the control strategies is how (or if) uncertainty in the forecast is incorporated into the control objective.



We demonstrate that an AWE system with remote sensing equipment can achieve a high level of performance using the most recent measurement to inform control decisions.
As the amount of information available to characterize wind speed profiles decreases, forecast uncertainty increases and performance declines.

Both results are related to the quality of the wind speed forecast, raising the question: Can a high-fidelity statistical model improve performance and/or close the gap in performance between different sensor configurations?
Our work underscores the need for further research exploring statistical methods for characterizing vertical wind speed profiles.
\section*{Acknowledgements}
This work was supported by the National Science Foundation under Award 1437296.
The authors would like to thank Cristina Archer at the University of Delaware for providing the wind speed data used in this study.

\bibliographystyle{IEEEtran}
\bibliography{refs}

\begin{thebibliography}{10}
\providecommand{\url}[1]{#1}
\csname url@rmstyle\endcsname
\providecommand{\newblock}{\relax}
\providecommand{\bibinfo}[2]{#2}
\providecommand\BIBentrySTDinterwordspacing{\spaceskip=0pt\relax}
\providecommand\BIBentryALTinterwordstretchfactor{4}
\providecommand\BIBentryALTinterwordspacing{\spaceskip=\fontdimen2\font plus
\BIBentryALTinterwordstretchfactor\fontdimen3\font minus
  \fontdimen4\font\relax}
\providecommand\BIBforeignlanguage[2]{{%
\expandafter\ifx\csname l@#1\endcsname\relax
\typeout{** WARNING: IEEEtran.bst: No hyphenation pattern has been}%
\typeout{** loaded for the language `#1'. Using the pattern for}%
\typeout{** the default language instead.}%
\else
\language=\csname l@#1\endcsname
\fi
#2}}

\bibitem{Diehl2013}
\BIBentryALTinterwordspacing
M.~Diehl, \emph{Airborne Wind Energy: Basic Concepts and Physical
  Foundations}.\hskip 1em plus 0.5em minus 0.4em\relax Berlin, Heidelberg:
  Springer Berlin Heidelberg, 2013, pp. 3--22. [Online]. Available:
  \url{https://doi.org/10.1007/978-3-642-39965-7\_1}
\BIBentrySTDinterwordspacing

\bibitem{altaeros2018}
\BIBentryALTinterwordspacing
Website, accessed: Sept 24, 2018. [Online]. Available:
  \url{http://www.altaeros.com/}
\BIBentrySTDinterwordspacing

\bibitem{skysails2018}
\BIBentryALTinterwordspacing
Website, accessed: Sept 24, 2018. [Online]. Available:
  \url{https://www.skysails.info/}
\BIBentrySTDinterwordspacing

\bibitem{ampyx2018}
\BIBentryALTinterwordspacing
Website, accessed: Sept 24, 2018. [Online]. Available:
  \url{http://www.ampyxpower.com/}
\BIBentrySTDinterwordspacing

\bibitem{archer2018}
\BIBentryALTinterwordspacing
C.~L. Archer, ``Wind profiler at cape henlopen,'' Website, accessed: Sept 23,
  2018. [Online]. Available:
  \url{https://www.ceoe.udel.edu/our-people/profiles/carcher/fsmw}
\BIBentrySTDinterwordspacing

\bibitem{loyd1980}
M.~L. Loyd, ``Crosswind kite power,'' \emph{J. Energy}, vol.~2, no. 80-4075,
  1980.

\bibitem{fagiano2014}
L.~Fagiano, A.~U. Zgraggen, M.~Morari, and M.~Khammash, ``Automatic crosswind
  flight of tethered wings for airborne wind energy: Modeling, control design,
  and experimental results,'' \emph{IEEE Transactions on Control Systems
  Technology}, vol.~22, no.~4, pp. 1433--1447, July 2014.

\bibitem{vermillion2018}
P.~Nikpoorparizi, N.~Deodhar, and C.~Vermillion, ``Modeling, control design,
  and combined plant/controller optimization for an energy-harvesting tethered
  wing,'' \emph{IEEE Transactions on Control Systems Technology}, vol.~26,
  no.~4, pp. 1157--1169, July 2018.

\bibitem{archer2014}
\BIBentryALTinterwordspacing
C.~L. Archer, L.~D. Monache, and D.~L. Rife, ``Airborne wind energy: Optimal
  locations and variability,'' \emph{Renewable Energy}, vol.~64, pp. 180 --
  186, 2014. [Online]. Available:
  \url{http://www.sciencedirect.com/science/article/pii/S0960148113005752}
\BIBentrySTDinterwordspacing

\bibitem{bafandeh2016}
A.~Bafandeh and C.~Vermillion, ``Real-time altitude optimization of airborne
  wind energy systems using lyapunov-based switched extremum seeking control,''
  in \emph{2016 American Control Conference (ACC)}, July 2016.

\bibitem{baheri2017}
\BIBentryALTinterwordspacing
A.~Baheri, S.~Bin-Karim, A.~Bafandeh, and C.~Vermillion, ``Real-time control
  using bayesian optimization: A case study in airborne wind energy systems,''
  \emph{Control Engineering Practice}, vol.~69, pp. 131 -- 140, 2017. [Online].
  Available:
  \url{http://www.sciencedirect.com/science/article/pii/S0967066117302101}
\BIBentrySTDinterwordspacing

\bibitem{brochu2010}
\BIBentryALTinterwordspacing
E.~Brochu, V.~M. Cora, and N.~de~Freitas, ``A tutorial on bayesian optimization
  of expensive cost functions, with application to active user modeling and
  hierarchical reinforcement learning,'' \emph{CoRR}, vol. abs/1012.2599, 2010.
  [Online]. Available: \url{http://arxiv.org/abs/1012.2599}
\BIBentrySTDinterwordspacing

\bibitem{soman2010}
S.~S. Soman, H.~Zareipour, O.~Malik, and P.~Mandal, ``A review of wind power
  and wind speed forecasting methods with different time horizons,'' in
  \emph{North American Power Symposium 2010}, Sept 2010, pp. 1--8.

\end{thebibliography}

\end{document}